\documentclass[onecolumn, aps, prb, longbibliography]
{revtex4-1}

\usepackage{graphicx}
\usepackage{caption}
\usepackage{braket}
\usepackage{color}

\begin{document}

\title{Hidden, entangled and resonating order}

\author{Gabriel Aeppli}
\affiliation{Laboratory for Solid State Physics, ETH Zurich, CH-8093 Z\"urich, Switzerland\\
Institut de Physique, EPFL, CH-1015 Lausanne, Switzerland\\
Paul Scherrer Institut,  CH-5232 Villigen, Switzerland}
\author{Alexander~V.\ Balatsky}
\affiliation{Nordita, Roslagstullsbacken 23, SE-106 91 Stockholm, Sweden\\
Dept.\ of Physics,
University of Connecticut,
Storrs, CT 06286, USA}
\author{Henrik~M.\ R\o{}nnow}
\affiliation{Laboratory for Quantum Magnetism, Institute of Physics, EPFL, CH-1015 Lausanne, Switzerland}
\author{Nicola~A.\ Spaldin}
\affiliation{Materials Theory, ETH Zurich, CH-8093 Z\"{u}rich, Switzerland}
\email[]{nicola.spaldin@mat.ethz.ch}

\begin{abstract}
In condensed matter systems, the atoms, electrons or spins can sometimes arrange themselves in ways that result in unexpected properties but that cannot be detected by conventional experimental probes. Several historical and contemporary examples of such hidden orders are known and more are awaiting discovery, perhaps in the form of more complex composite, entangled or dynamical hidden orders.

\vspace*{1cm}

\color{red}{This is the Author Accepted Manuscript of an article published in {\it Nature Reviews Materials} {\bf 5}, 477–479 (2020), \copyright 2020, Springer Nature Limited. 

\vspace*{0.5cm}
The final authenticated version is available online at: https://doi.org/10.1038/ s41578-020-0207-z
}
\end{abstract}

\date{June 04, 2020}

\maketitle

\subsection*{An early example of hidden order}
Almost one hundred years ago, an unexpected experimental observation was made: in MnO, a cusp in the specific heat as a function of temperature \cite{Millar:1928}, indicative of a phase transition, was found to coincide with a cusp in the magnetic susceptibility \cite{Tyler:1933}, suggesting that the phase transition had a magnetic origin. Although such a combination was already known to occur at transitions to the ordered magnetic states in ferromagnetic materials, it was inexplicable in MnO, which has no net magnetic moment at low temperature. 

A convincing explanation was soon proposed by N\'eel \cite{Neel:1936}, who showed theoretically that the observed behavior was consistent with  hidden antiferromagnetic order, corresponding to an  antiparallel alignment of neighbouring Mn magnetic moments. This was a bold suggestion for the time, because the superexchange description of chemical bonding that justifies antiparallel alignment of magnetic moments was not yet developed. However, it proved to be correct and earned N\'eel a Nobel Prize after a probe was identified --- neutron diffraction --- that was able to directly verify the antiferromagnetic ordering \cite{Shull/Smart:1949}. Amusingly, N\'eel  made the statement in his Nobel lecture that while ``Today, an extremely large number of antiferromagnetic substances are known . . . They are extremely interesting from the theoretical viewpoint, but do not appear to have any practical applications.''  In fact, antiferromagnetic materials underpin multi-billion dollar industries as the exchange bias component in magnetic sensors and are promising for possible future spintronic devices.

\subsection*{Modern hidden orders}
In recent decades, a number of other kinds of hidden order have been proposed, which are of fundamental interest and in some cases also of possible technological importance. One example is emphanisis, the emergence of correlated local electric dipoles upon warming \cite{Bozin_et_al:2010}, first identified in the semiconductor PbTe (Fig. 1a). The Pb$^{2+}$--Te$^{2-}$  dipoles are hidden in the sense that they don't line up to produce a macroscopic ferroelectric polarization, and their order is only revealed using sensitive probes of the local structure. Emphanisis is important for the favorable thermoelectric behavior of PbTe, because the hidden locally ordered dipoles suppress the thermal conductivity and enhance the Seebeck coefficient, and might also be relevant for PbTe's exotic superconductivity. 

Another example is the long-standing puzzle of the hidden order underlying the  observed low-temperature (17.5K) phase transition in the heavy-fermion compound URu$_2$Si$_2$ \cite{Mydosh/Oppeneer:2014}. The phase transition indicates an unconventional electronic and magnetic hidden order  with an origin that is still a matter of discussion, with leading proposals including an entangled, so-called hastatic, order between the localized $5f$ spins on the uranium ions and the mobile conduction electrons, as well as high-order charge multipoles, ranging from quadrupoles through octupoles and hexadecapoles all the way to triakontadipoles (high-order charge multipoles of rank 5).  Such charge multipoles are indeed established in other actinide-based systems; for example, antiferroquadrupolar ordering of the $5f$ quadrupoles on the uranium ions in UPd$_3$ (Fig.  1b) has been demonstrated using X-ray resonant scattering \cite{McMorrow_et_al:2001}. 
A hidden electronic nematic order is also a candidate. Named by analogy to the preferred direction adopted by molecules in nematic liquid crystals, in electron nematics the electrons are believed to generate a spontaneous anisotropy that is not reflected in the symmetry of the crystal lattice. In addition to its possible relevance for URu$_2$Si$_2$, electronic nematicity has been proposed to account for the anisotropic conductivity in materials ranging from the pnictide superconductors to the strontium ruthenates. The latter material system is a good example of how some suggested hidden orders have been abandoned as they have been excluded by improved characterization techniques; in this case neutron scattering has revealed incommensurate Bragg peaks in Sr$_3$Ru$_2$O$_7$ consistent with a  spin density wave, providing a more straightforward explanation for the material's behaviour\cite{Lester_et_al:2015}. As a last well-established example, we mention the topological quantum hidden order associated with the fractional quantum Hall effect, which is particularly  interesting because it is accompanied by quantum entanglement of the electrons. Such quantum-entangled hidden order also manifests in various paradigmatic magnetic models such as the Kitaev model and the Haldane state in 1D spin chains. 

\subsection*{ Searching for new kinds of hidden order}
The discovery of new physics, combined with the allure of novel device applications, motivate a systematic search for additional kinds of hidden orders. Based on the premise that most straightforward kinds of order have already been discovered, we propose composite orders as an appealing area for study. In this scenario, a product of two orders described by order parameters $A$ and $B$ could have a non-zero value, even if neither $A$ nor $B$ are ordered individually. Mathematically, 
\[ \langle A \otimes B \rangle \neq 0  \textrm{  even though  } \langle A \rangle  = 0 \textrm{   and  } \langle B \rangle = 0  \quad .\] Such hidden composite order could be classical, or based on quantum entangled states,
$\langle \Psi_A \otimes \Psi_B \rangle \neq 0$, or could be a dynamical order, $\langle A(t) \otimes B(t') \rangle \neq 0$, for which no stationary description exists, suggesting a hierarchy of possible new hidden, entangled and resonating or dynamical orders. 

\paragraph*{Hidden magnetoelectric order.}
One promising direction is motivated by multiferroic materials, which are simultaneously ferromagnetic and ferroelectric (Fig. 1, leftmost panel); that is, the order parameters describing the magnetic moment, $\mu$, and the electric dipole, $d$, are non-zero: 
\[ \langle \mu \rangle \neq 0  \textrm{    and  } \langle d \rangle \neq 0  \quad .\]
Instead of this revealed composite order, one can imagine a scenario in which both magnetic and ferroelectric dipole orders are zero but their product is not (Fig. 1c), that is 
\[ \langle \mu \otimes d \rangle \neq 0  \textrm{  even though  } \langle \mu \rangle  = 0 \textrm{   and  } \langle d \rangle = 0  \quad .\]
Such a material would have no net magnetization or ferroelectric polarization and so would not have the usual signatures in the magnetic and dielectric susceptibilities associated with ferromagnets and
 ferroelectrics, respectively. Instead it would have a new hidden order with different, perhaps more complex, susceptibilities associated with its coupled breaking of both space-inversion and time-reversal symmetry. 

Such composite magnetoelectric orders might already be hiding in plain sight, for example in the high-temperature cuprate superconductors, that have an intricate phase diagram that is a rich source of exotic physics, including phases with as yet poorly understood kinds of order. In particular, the so-called pseudo-gap phase, which occurs above the superconducting dome in under-doped cuprates, has indications that it breaks both time-reversal and space-inversion symmetries, although it has no net magnetism and polarization. A leading proposal of orbital currents, in which each unit cell hosts a pair of electron-current loops flowing in opposite directions, is consistent with the observed time-reversal and space-inversion symmetry breaking \cite{Varma:2006} (Fig. 1d). A complementary suggestion of dynamically ordered quasistatic magnetoelectric multipoles, in which both spins and electric dipoles fluctuate but their ordered product state is stabilized through spin--phonon coupling, yields the same combination of symmetry breakings (Fig. 1e). 

\paragraph*{Hidden dynamical order.}
A dynamical order originating from fluctuations is of course not restricted to magnetic and electric dipoles, and we anticipate it could be quite prevalent. A possible example is the composite odd-frequency-paired boson condensate between a conventional Cooper pair, $c_\alpha(\textbf{r},\tau)c_\beta(\textbf{r},\tau)$, where $\alpha$ and $\beta$ are spin indices, and a neutral boson spin field,  $\textbf{S}(\textbf{r})$, suggested in dynamic superconducting states (Fig. 1f). Such a coupling could occur even when neither the Cooper pairs nor the spins individually order, that is
\[
    \langle c_\alpha(\textbf{r},\tau)c_\beta(\textbf{r},\tau)\rangle = 0 \textrm{   and  } \langle \textbf{S}(\textbf{r})\rangle = 0 \quad ,
\]
yet
\[
     \langle c_\alpha(\textbf{r},\tau)c_\beta(\textbf{r},\tau) \varepsilon_{\alpha \beta}\bf{S}(\textbf{r}) \rangle \neq 0 
\]
 ($\varepsilon_{\alpha \beta}$ is the antisymmetric tensor). This example is particularly intriguing as it demonstrates the reversal of the conventional relationship between spin singlet ($S=0$) $s$-wave and spin triplet ($S = 1$) $p$-wave pairing channels, allowing instead $S = 1$ $s$-wave and $S = 0$ $p$-wave pairs.
Heterostructures that combine magnetic and superconducting films are prime candidates for hosting such composite entangled superconducting orders, because strong pairing fluctuations in the magnetic background should induce composite states at the interfaces. 

Dynamics can also be exploited to drive a system out of equilibrium using strong, short laser pulses, and stabilize ordered states that do not exist under equilibrium conditions. An example is the hidden electronic state induced in the trigonal phase of TaS$_2$ (Fig. 1g), which has modified resistivity, excitation spectra and optical reflectivity compared to the ground state \cite{Stojchevska_et_al:2014}.

A search for new types of hidden orders and materials that host them is particularly timely. Electronic structure methods, such as Berry phase theories that allow direct calculation of hidden multipoles, are now available, and high-throughput or machine-learning techniques could prove valuable in identifying new materials classes with targeted properties. New state-of-the art tools, such as free-electron lasers and polarized neutron analysers, combined with progress in closing the terahertz gap, enhance the possibilities of detection.  Finally, the identification of non-trivial entangled quantum orders over mesoscopic or macroscopic distances could enable the exploitation in solid-state devices of the `spooky action at a distance', as Einstein called quantum entanglement.

\subsection*{ Acknowledgements}
This work has received funding from the European Research Council (ERC) under
the European Union’s Horizon 2020 research and innovation programme project HERO grant agreement No.\ 810451.

\subsection*{ Author contributions}

All the authors contributed equally to the preparation of the manuscript.

\subsection*{ Competing interests}

The authors declare no competing interests.

    \centering
    \includegraphics[angle=270,width=\linewidth]{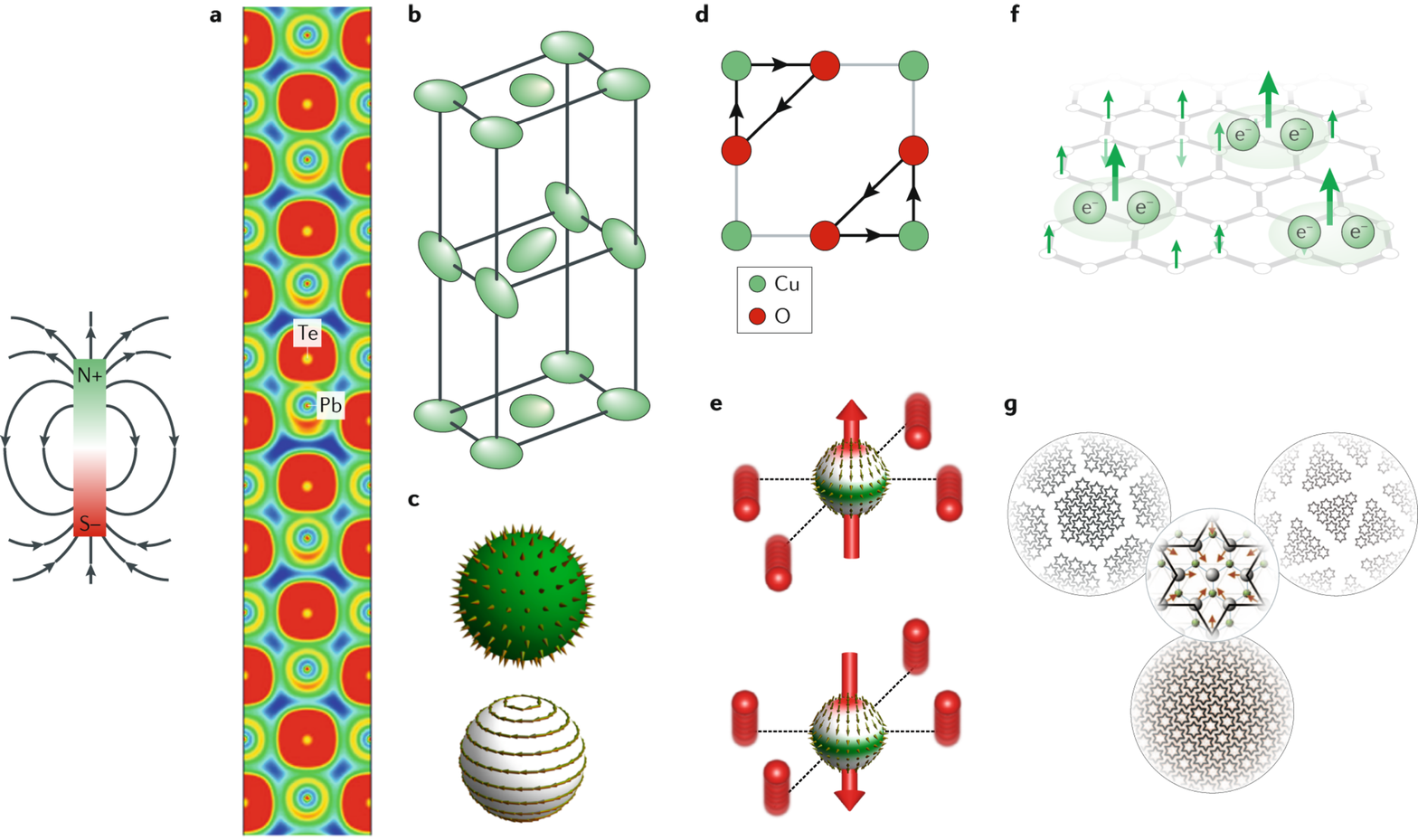}

{\bf Figure 1| Examples of hidden orders.} Conventional orders such as ferroelectricity and ferromagnetism have been well established for centuries and are widely exploited in applications. Composite versions of these orders, such as multiferroicity, are more recent but also well characterized. Interesting new research directions and potential technologies are suggested by a wide range of novel hidden composite, dynamic and entangled orders that build on these conventional states. {\bf a)} Hidden electric dipoles. Hidden correlated local electric dipoles associated with the large thermoelectric response of PbTe. The color indicates the size of the electron localization function, with high values in red and low values in blue. The central Pb ion is displaced from the center of its surrounding Te ions, making a local dipole that propagates to neighboring unit cells. {\bf b)} and {\bf c)} Hidden higher-order charge or magnetic multipoles. {\bf b)} Proposed hidden antiferroquadrupolar order in UPd$_3$ with the uranium 5f quadrupole moments shown as ellipsoids. {\bf c)} Magnetoelectric multipoles occur in inversion-symmetry-broken magnets and cause a magnetic response to an electric field. The cartoons show the magnetoelectric monopole, for which $\langle \mu \cdot d \rangle \neq 0  $, and the magnetoelectric toroidal moment, for which $\langle \mu \otimes d \rangle \neq 0  \textrm.$  The arrows indicate the orientation of the local magnetization. {\bf d)} and {\bf e)} Proposed hidden orders in high-temperature superconductors. {\bf d)} Proposed orbital currents in the pseudogap phase of the cuprates. The arrows indicate the proposed hidden circulating orbital currents that break both time-reversal and space-inversion symmetries. {\bf e)} Coupling between spin fluctuations (red arrows) and phonon fluctuations (red spheres) could lead to a conserved magnetoelectric multipole, in which the product of spin and position does not fluctuate. In this case the magnetoelectric quadrupole keeps the same sign when the spin reverses sign, because the reversal of the spin is accompanied by reversal of the optical phonon. {\bf f)} and {\bf g)} Hidden dynamical (or dynamically induced) orders . {\bf f)} Composites of Cooper pairs (blue spheres and large arrows) and (spin or lattice) bosons  (small arrows) are proposed to lead to odd-frequency pairing in superconductors. {\bf g)} Polarons in TaS$_2$ cause star-shaped lattice distortions (center) that can order in various patterns that were until recently hidden, including one (top right) that is dynamically induced.Part {\bf a)} adapted from B. Sangiorgio et al., Phys. Rev. Materials 2, 085402 (2018). Part {\bf b)} [adapted from ref. 7]. Part {\bf c)} [adapted from N. A. Spaldin et al., Phys. Rev. B 88, 094429 (2013)]. Part {\bf d)} [adapted from ref. 9]. Part {\bf e)} adapted from [M. Fechner et al., Phys. Rev. B 93, 174419 (2016)]. Part {\bf f)} [adapted from J. Linder and A. V. Balatsky, Rev. Mod. Phys. 91, 045005 (2019)]. Part {\bf g)} adapted from Ref. 10, AAAS.

\end{document}